# Dominant misconceptions and alluvial flows between Engineering and Physical Science students

Anna Chrysostomou[1], Alan S. Cornell[2], and Wade Naylor[3]

[1]Laboratoire de Physique Théorique et Hautes Énergies - LPTHE, Sorbonne Université, CNRS, 4 Place Jussieu, 75005 Paris, France

[2]Department of Physics, University of Johannesburg, PO Box 524, Auckland Park 2006, South Africa

[3]National School of Education, Faculty of Education and Arts, Australian Catholic University, Brisbane, Australia

Abstract

In this article we assess the comprehension of physics concepts by Physical Science and Engineering students enrolled in their first semester at the University of Johannesburg (UJ), South Africa (2022). We employ different graphical measures to explore similarities and differences using the results of both pre- and post-test data from the Force Concept Inventory assessment tool, from which we calculate dominant misconceptions (DMs) and gains. We also use alluvial diagrams to track the choices made by these two groups of students from pre- to post-test stages. In our analysis, we find that DM results indicate that participating Engineering students outperformed Physical Science students on average, however, the same types of normalised DMs persist at the post-test level. We call these DMs "persistent misconceptions." This is very useful when tracking persistent misconceptions, where when using repeated measures and alluvial diagrams with smaller groups of students, we find that Physical Science students tend to make more chaotic choices.

---

Wade Naylor https://orcid.org/0000-0003-2133-7924 E-mail: wade.naylor@acu.edu.au (author for correspondence)
Alan S. Cornell https://orcid.org/0000-0003-1896-4628 E-mail: acornell@uj.ac.za
Anna Chrysostomou https://orcid.org/0000-0002-7720-9369 E-mail: chrysostomou@lpthe.jussieu.fr

The authors have followed the Contributor Roles Taxonomy:
Anna Chrysostomou - Conceptualisation, Methodology, Data Curation, Writing - review & editing;
Alan Cornell - Conceptualisation, Methodology Writing - original draft, Funding acquisition;
Wade Naylor - Conceptualisation, Methodology, Formal analysis, Investigation, Visualisation, Project Administration, Writing - review & editing, Funding acquisition.

## Introduction

Given the foundational placement of classical mechanics in the development of conceptual understanding in physics among first-year university students, it is an important area of study within physics education research (PER). Recall that physics is categorised as a "hierarchical knowledge structure" (Bernstein, 1999, 2000): it is a coherent framework constructed using precisely defined concepts and the relationships between them (Cornell & Padayachee, 2021), which integrates knowledge at lower levels to expand into increasingly complex and complicated ideas. In this metaphor, classical mechanics serves as the foundation stone for the sprawling body of knowledge that is physics. As such, research into how students engage with misconceptions in their classical mechanics course(s) is essential in order to help to improve the teaching of physics on deeper conceptual levels. Such research has led to the development of various concept inventories (CIs), known also as research-based assessments (RBAs), that seek to assess students' understanding; for examples, see Le et al. (2024), Madsen et al. (2017b), and Santoso et al. (2024) in the context of PER. Note that in the work of Madsen et al. (2017a, 2017b), the authors have explored a whole suite of physics CIs, also including a level of research validation for each CI and how best to deploy them for evaluating student conceptual understanding.

The goal for CIs is to obtain insights into students' conceptual understanding of key physics concepts, with a hope that conceptual gains will increase with instruction.[1] The Force Concept Inventory (FCI) of Hestenes et al. (1992) is a widely used CI which seeks to assess concepts in Newtonian mechanics. The test itself consists of 30 multiple-choice questions, each of which has four possible responses; only one answer per question is correct. Notably, the incorrect answers are crafted based on "commonsense beliefs" that students often develop through intuitive, yet incorrect, interpretations of everyday experiences (Hestenes et al., 1992). Thus, the FCI serves as a "diagnostic tool" that helps educators identify specific misconceptions, after which they can tailor their teaching strategies to address these misunderstandings effectively. The FCI has been used to identify common misconceptions by comparing pre- and post-test results (that is, comparing a "pre-test" taken before students' first Newtonian mechanics course, and a "post-test" taken after the course), where educators can subsequently gauge the effectiveness of their instruction and focus on areas that may need revisiting, for example see Carleschi et al. (2022) and Chrysostomou et al. (2024).

It should be further noted that the work of Santoso et al. (2024), who performed a meta-analysis of the role of gender based on a study of $n = 38$ FCI papers, is another example of how RBAs can be used in PER. One of the ultimate goals of any kind of RBA is that the findings from such studies should affect and update teaching practice to help students in physics courses achieve a better conceptual understanding of physics concepts, regardless of their background. Rather than simply compare pre- and post-test responses on a year-to-year basis, this analysis can be done right after the pre-test (as suggested in Chrysostomou et al. (2024)) by studying the "dominant misconception" (DM) − the most common incorrect response and the underlying rationale for choosing it − for each question of a CI. We shall explain DMs in more detail in the Methodology Section.

[1]The issue of how gain is actually evaluated is non-trivial, but there are mathematically consistent ways to define them, for example see Nitta and Aiba (2019) and Naylor et al. (2024).

Another interesting aspect of RBA is the comparison of conceptual understanding between different groups or cohorts of students, such as Engineering and Physical Science students[2] in their first year of university studies. Given that the FCI has been one of the most researched and resourced CIs to date (Hake, 1998; Hestenes et al., 1992) (where we acknowledge that there has also been work done on Engineering CIs, for example see Jorion et al. (2015) and Naylor et al. (2024)), there appears to be a gap in the literature, where no comparison has been done between Engineering and Physical Science students using the FCI. The absence of such a comparison is surprising, as both cohorts of students study mechanics in their first semester of their first year at university − attending the same lectures and/or tutorials, in some contexts, or classes teaching similar concepts and skills, in parallel. Note that filling this deficit in the literature offers an opportunity to look at possible differences between each group, and how their different career goals may influence their approach to and comprehension of physics concepts. As we shall discuss later, our results suggest that such a discrepancy does indeed exist and can be observed at the "pre-test level" (see Table 4). By studying these conceptual differences in each group directly after the pre-test, lecturers can adjust their teaching strategies to manage the conceptual needs of each group. In this way, we highlight that the FCI can be used as a diagnostic tool to assess students' understanding of key concepts in mechanics and to identify common misconceptions (Hestenes et al., 1992).

For example, in previous studies involving the FCI, data were collected from large cohorts of diverse students at the University of Johannesburg (UJ) (Carleschi et al., 2022; Chrysostomou et al., 2024), where we considered the results collectively rather than in separate course-based cohorts. In re-analysing the 2022 FCI pre-test data, it shall be revealed in this work that there were differences in misconceptions between the Engineering cohort and a Physical Science cohort. As such, looking at these results in separate cohorts provides an opportunity to explore how these groups might possibly engage with and develop their understanding of physics differently. From previous studies, it has also been demonstrated that misconceptions in physics are prevalent and can impact on a student's ability to grasp or develop particular concepts (Carleschi et al., 2022; Chrysostomou et al., 2024). With this in mind, this paper aims to extend previous work by focusing specifically on the differences in conceptual understanding between Engineering and Physical Science students at UJ during their first year of study.

## Physical Science and Engineering perspectives

Whilst both Physical Science and Engineering students are required to have strong problem-solving skills in order to enter their study programmes, there are distinct differences between these two directions of study. Broadly speaking, physics-based programmes focus on understanding and explaining the fundamental laws and theories that govern observed phenomena, whereas engineering programmes adopt a more practical and applied approach.

For this reason, Physical Science students are expected to have a strong theoretical foundation in scientific principles and methodologies. This includes the ability to formulate hypotheses, design experiments, and analyse data. Critical thinking and attention to detail are paramount, as scientific research often involves meticulous observation and interpretation of results (National Research Council, 2003). While problem-solving skills are also important for Physical Science students, there is often a greater emphasis on theoretical understanding and research than would be the case for engineering students. As such, Physical Science students must be adept at navigating scientific literature, understanding complex concepts, and contributing to ongoing research discussions. Effective communication skills are also crucial, particularly for writing scientific papers and presenting findings (Snow & Dibner, 2016). While teamwork may be needed in some disciplines, this is not as strongly enforced as in engineering classes in many institutions.

[2] By "Physical Science students", we mean students enrolled in a double-major Bachelor of Science (BSc.) programme who have chosen Physics and one other Physical Science subject (e.g. Mathematics, Chemistry, Computer Science, Applied Mathematics, etc.) for their majors (see Table 1).

For Engineering students, their strong analytical and problem-solving skills are crucial for designing and optimising systems, structures, and processes. According to Felder (2002), successful Engineering students often exhibit strengths in logical reasoning, spatial visualisation, and quantitative analyses. Additionally, they must be proficient in applying theoretical knowledge to practical problems, a skill that is developed through hands-on laboratory work and project-based learning. Quite interestingly, communication skills are also essential for Engineering students, as they must be able to convey complex technical information clearly and effectively, both in writing and verbally. Collaboration is another critical skill, given that engineering projects often involve teamwork across various disciplines (Brophy et al., 2008). These practices are often reinforced by the group work component in their laboratory classes.

Another major distinction to consider is the university entry requirements for each programme. Engineering programmes are typically more selective in their admissions processes, reflecting the rigorous nature of the curriculum which is demanded by the national accrediting body – in South Africa, this is the Engineering Council of South Africa (ECSA). To maximise student success and manage limited university space, engineering faculties throughout the world require strong backgrounds in mathematical and physical sciences. For example, applicants in the United States typically need advanced high school coursework in Calculus, Physics, and Chemistry (Brophy et al., 2008), while universities in the United Kingdom often demand high A-level grades in Mathematics and Physics ("UCAS Undergraduate entry requirements," n.d.). In several Asian countries, entrance to engineering programmes is extremely competitive, with exams like China's Gaokao (Wang et al., 2022) and India's Joint Entrance Examinations (Advanced) ("JEE (Advanced)," n.d.) assessing a wide range of subjects but emphasising maths and science proficiency. As far as we are aware the only other work that looks at engineering cohorts is that by Martín-Blas et al. (2010); there was no comparison to a Physical Science cohort.

Physics programmes, on the other hand, are also rigorous but generally have more flexible entry requirements due to their broader range of majors, including Biology and Chemistry. Applicants to Physical Science degrees need strong backgrounds in relevant sciences, but Physics programmes may not require the same level of specialisation as in Engineering. In the United States, Physics programmes require strong performance in Physics and Mathematics, but not always as specialised as for Engineering. In the United Kingdom, Physics programmes require A-levels in one science subject, with varying additional requirements. For example, a Biology programme might require A-levels in Biology and another science, while a Physics programme might require A-levels in Physics and Mathematics ("UCAS Undergraduate entry requirements," n.d.).

In the UJ context, the different expectations placed on students can be seen explicitly in Table 1, where we summarise the entry requirements for three programmes. Specifically, we present the entry-level requirements for English, Mathematics, and Physical Sciences (the subject covering physics and chemistry modules in South African high schools). We also include the South African Admission Points Score (APS): a converted average corresponding to a student's final year (Matriculation) mark, used by South African higher education institutions to evaluate a student's eligibility for admission. The APS is calculated by assigning point values to a student's final Matriculation marks for their top six subjects (excluding Life Orientation), with each subject's percentage falling into a specific bracket that corresponds to a point value (e.g. 7 for 80-100%, 6 for 70-79%, etc.), and then summing these points (Wilbesi, 2022). The highest possible APS is 42.

**Table 1**

*Minimum entry requirements at UJ. Each row corresponds to the degree: "BEng." refers to a Bachelor in Electrical and Electronic Engineering , "BPhys." is our shorthand for the extended BSc. in Physical Sciences (Physics and Mathematics) programme, and "BPhys. (trad.)" is the "traditional" BSc. in Physical Sciences (Physics and Mathematics) , respectively. The length of the degree course is written in brackets in the first column; for English, Maths, and Physical Science, we list the minimum APS requirement and its corresponding minimum percent in brackets (see text for details).*

| Course / Length (yr.) | APS | English | Maths | Physical Science |
|---|---|---|---|---|
| BEng. (4 yrs) | 32 | 5 (60%) | 5 (60%) | 5 (60%) |
| BPhys. (4 yrs) | 26 | 4 (50%) | 5 (60%) | 4 (50%) |
| BPhys. (trad.) (3 yrs) | 31 | 5 (60%) | 6 (70%) | 5 (60%) |

In Table 1, we include the four-year Bachelor of Engineering (BEng.) programme, the traditional three-year BSc. in Physical Sciences (BPhys. (trad.)) programme, and the four-year BSc. in Physical Sciences Extended (BPhys.) programme.[3] Both the BPhys. (trad.) and the BPhys. programmes have two majors (for example, Physics and Mathematics) with additional credits from other modules (for example, Computer Science or Applied Mathematics); unlike the "regular" BPhys. (trad.) programme, the BPhys. is a four-year programme where the first three semesters bridge between the end of high school and the beginning of the second semester of the conventional three-year BPhys. (trad.) programme. Specifically, within the first year of the course, mandatory additional classes are included – "Computer Competence" (one-semester course) and "Language for Science" (full-year course) – in order to fill the gaps in students' high school education. The first year focuses on the majors (e.g. Physics and Mathematics); from the second year, "minors" are included in the form of additional supporting modules e.g. Applied Maths or Computer Science, etc. (see section  for details).

[3]Note that while "BEng." is a standard abbreviation used in official UJ documentation Faculty of Engineering and the Built Environment, 2022; Faculty of Science, 2022, "BPhys." and "BPhys. (trad.)" represent a shorthand used in this article.

To understand how the entry requirements differ between Physical Science majors and Engineering students, we compare the entry requirements for the BEng. programme (Faculty of Engineering and the Built Environment, 2022), the BPhys. programme (Faculty of Science, 2022), and the BPhys. (trad.) programme (Faculty of Science, 2022), respectively. The BEng. programme requires the highest APS (32) and a consistent level of performance across English, Mathematics, and Physical Science, with a minimum of 60% in each. The BPhys. programme has the lowest minimum APS (26) and lower requirements in both English and Physical Science (50%). However, like the BEng. programme, the BPhys. programme has a 60% entry requirement for Mathematics. In comparison, the traditional BSc. Phys. Sci. programme requires a high APS (31) and the highest level in Mathematics (70%) among the three. As such, in keeping with the global trend, the BEng. course has the highest barrier to entry due to the highest overall APS requirement and the consistent performance required across subjects. BPhys. (trad.) also has challenging requirements, particularly in Mathematics, but the slightly lower APS makes it slightly easier to access than the BEng. course.

In this study, we focus on two first-year UJ cohorts: first-year students pursuing their first semester of physics in the BEng. programme and the BPhys. programme (see the first two rows in Table 1), where we are motivated to compare these two cohorts due to their similar entry requirements for Mathematics and Physical Sciences as well as their population size (we shall elaborate on our choice of cohorts and the structure of their courses in the *Participants* subsection of the Methodology Section). It may also be worth mentioning that in the literature (Marginson et al., 2013) students in general need much higher maths and science scores to succeed in Engineering subjects, where an APS of 32 may well still not fully prepare students on such Engineering courses. As such, we wish to investigate the extent to which this is reflected in the students' FCI scores through the lens of DMs. In particular, we seek to understand if the misconceptions present at the pre-test level differ between these two cohorts, and then persist or not at the post-test stage.

Since the FCI measures not only conceptual knowledge, but also the ability to apply this knowledge in various contexts, it is a useful tool to evaluate the academic performance of both Engineering and Physical Science students. Furthermore, the FCI is one particular RBA that has been intensely studied in its conceptual structure (see Tables A1 and A2), as well as in terms of an item response theory (IRT) structure (i.e. difficult vs. easy questions and student ability (Wang & Bao, 2010)). As stated earlier, this can include a question-by-question breakdown of correct and incorrect response rates, which can give a snapshot of the types of questions that each cohort struggles with (see Figures 1, 2) and the corresponding DMs. Moreover, we can also analyse how each answer choice for a given questions evolves from pre- to post-test using alluvial diagrams. To elaborate on this, in the next section, we will detail the visualisation tool we will use in this studay and its value within the context of PER.

Within the context of our study and our main objectives, we can summarise the direction of our investigation through two research questions, where the use of alluvial diagrams will greatly highlight the development of ideas and DMs especially in the case of students who do not make an attempt on a question (a non-attempt response often not having been captured in previous studies). We introduce these in the next section.

### Research questions

Our comparison of the conceptual understanding of Engineering and Physical Science students shall be guided by the following two research questions:

**RQ1–** To what extent do dominant misconceptions compare between Physical Science students in the extended programme (i.e. students pursuing a four-year BSc. in Physical Sciences, where one major is Physics; in other words, BPhys. students) and Engineering (i.e. BEng.) students at the University of Johannesburg?

**RQ2–** How do alluvial diagrams compare between Physical Science and Engineering students, in terms of their precise responses to FCI questions?

As already mentioned, the data for this study was collected during 2022 from a large, ethnically and culturally diverse cohort of first year Physical Science and Engineering students (see the Participants section for details). The results indicate that while Engineering students generally enter the courses with higher FCI scores (i.e. perform better on the pre-test, on average) compared to their Physical Science counterparts, the trajectory of their conceptual development presents unique patterns worth investigating.

In what follows, we will discuss the methodology used in our study, present our results and analyses, and conclude with recommendations for future work. Our goal is to provide insights into how different student cohorts develop their understanding of physics and to inform teaching strategies that can better address persistent misconceptions. This work aims to contribute to the broader discourse on PER and the effective teaching of mechanics to diverse student populations.

## Methodology – Gain, dominant misconceptions, and alluvials

To illustrate changes in students' understanding, we shall employ Hake gains, DMs, and alluvial diagrams. Hake gain is nothing but the repeated measures or "paired gain", which measures the improvement in individual students' scores from pre- to post-test stages, to quantify learning gains within and between the cohorts. From gains which require a wait time until the post-test, we discuss how to quantify DMs, which can be used right after the pre-test. As well as DMs, alluvial diagrams visualise the flow of students' choices on selected FCI questions from pre- to post-test stages and highlight the shifts in students' conceptual frameworks and the persistence of certain misconceptions over time.

### Hake gain

Traditionally, in pre-/post-test investigations we quantify the conceptual uptake of the participants using the "gain". Here, following the usual FCI approach, we employ the normalised gain $G$, formally defined as (Hake, 1998):

$$G = \frac{\langle \%S_f \rangle - \langle \%S_i \rangle}{100 - \langle \%S_i \rangle} \, , \tag{1}$$

where $\%S_f$ and $\%S_i$ are the final and initial FCI test scores, respectively. In Hake's seminal paper, covering 62 introductory courses and a total of 6542 students, he suggested the following classifications:

- high-gain courses: $G \geq 70\%$;
- medium-gain courses: $30\% \leq G < 70\%$;
- low-gain courses: $G < 30\%$.

It is interesting to note that in his survey, Hake found that all students participating in "traditional" lecture-based courses ($N = 2084$) scored in the low-gain category, with an average score of $\langle G \rangle = 0.236 \pm 0.04$sd. In contrast, 85% of courses involving "interactive engagement" fell into the medium-gain regions with only the remaining 15% falling into the low-g category (Hake, 1998). As a result of Hake's work, a gain of approximately 24% has since come to be accepted in the literature as a historic benchmark. This benchmark has since been widely referenced in PER to represent the typical effectiveness of conventional teaching methods in improving students' conceptual understanding of mechanics. For instance, the PhysPort website (McKagan et al., 2020) reports an average normalised gain of 22% for traditional lecture courses, corroborating Hake's findings.

We acknowledge, however, that there are limitations associated with this variable. An immediate challenge is that we need to wait for the post-test results to analyse the data. Educators, therefore, cannot use $G$ to improve their teaching strategies for the cohort they are in the midst of teaching; the results can only inform the following academic year. Furthermore, Eq. (1) only quantifies the difference between pre- and post-test, but fails to take into account the actual FCI test score. In other words, $G$ does not distinguish between a gain from, say, 5 to 10 or 25 to 30, from pre- to post-test.[4] Finally, even with large samples sizes repeated measure (paired) data is always smaller than the original sample size.

### Normalised dominant misconceptions

As described in the introduction, DMs in the FCI context represent a common incorrect response and the underlying rationale for choosing it. These were first described in the seminal work of Martín-Blas et al. (2010). To adjust for the frequency of misconceptions across populations of different sizes, we make use of the "normalised" DM (nDM). To explain this parameter, consider the FCI multiple choice options, of which there are five – $A, B, C, D, E$. There is also another option: not answering at all, $NA$. In which case the following equation can be used to calculate the %nDMs for a given question,

$$\%nDM = \frac{\%(\text{largest incorrect answer})}{100 - \%(\text{correct answer}) - \%(NA)} \times 100 \; , \tag{2}$$

where the subtraction of the $NA$ and correct answers serve as the normalising components. The advantages of incorporating the percentage of correct answer responses in the normalisation lies in the fact that DMs "drowned out" by correct answers are amplified; similarly,

[4]There are ways to avoid this issue using Rasch and IRT models (Naylor et al., 2024; Nitta & Aiba, 2019).

"subdominant misconceptions" can also be exposed. Furthermore, this normalisation allows for the identification of DMs even in the absence of paired data and/or even if only the pre-/post-test is performed. A threshold is chosen, such as 50% (Bani-Salameh, 2016a; Martín-Blas et al., 2010), where anything above this "cutoff" is deemed a DM. For example, suppose the correct answer to Q1 is C, but 51% of incorrect answers fall on A for Q1; then the (erroneous) "commonsense belief" used to arrive at A indicates a misconception.[5] We call this misconception a "dominant" one due to its prevalence in the tested population. This, of course, means that some question items do not have a DM (i.e. their incorrect answers lie below a certain % threshold), or that there could be more than one incorrect choice which is dominant, again above a certain % threshold. As further discussed by Bani-Salameh (2016a), the actual value for the threshold (the % where the incorrect answer is defined as dominant) can be shifted. Like in Martín-Blas et al. (2010), in this study we have also chosen 50% for the threshold for the two UJ cohorts.

Although we are motivated to compare the FCI results of BPhys. and BEng. students due to their large class sizes, it is difficult to find equivalently large sample sizes of "paired data". This means that robust statistical methods (used for smaller sample sizes) are often needed. For this reason, nDMs are particular useful in studying DMs even after assuming an overall gain in the correct answer choices from pre- to post-tests. As alluded to earlier, investigations based on DMs can be performed using data solely from either the pre-test or post-test results.

**Table 2**

*Selected Newtonian concepts in the force concept inventory (FCI).*

| Question No. | Newtonian Concept[a] |
|---|---|
| 4, 28 | Non-equal action-reaction pairs |
| 19 | Position/velocity/acceleration undiscriminated |
| 2 | Displacement time depends on the mass |
| 9 | Non-vectorial velocity composition |
| 14 | Ego-centred reference frame |

[a]The Newtonian concepts considered here come from the taxonomy described in Tables A1 and A2.

A further benefit of using DMs is that we can compare our results against those from other authors, to gain insight into the teaching practices of international cohorts. For example, PER in Spain (Martín-Blas et al., 2010) and the Kingdom of Saudi Arabia (KSA) (Bani-Salameh, 2016a, 2016b) include complete tables of data and their corresponding tested concepts can be found for all questions in those works. For example, Martín-Blas et al. (2010) found a given set of questions that had large DMs were well correlated between the two cohorts studied (Martín-Blas et al., 2010) and related to particular Newtonian concepts, see Table 2. These particular choices have been broadly classified based on the taxonomy of FCI questions, as first developed by Hestenes et al. (1992) and adapted by others (Bani-Salameh, 2016a; Bayraktar, 2009; Martín-Blas et al., 2010). We present them here, for completeness, with our own slight adaptations in Appendix A; see Tables A1 and A2. This motivates the question choices and ordering given in Table 6.

[5]As explained in the introduction, this assumption is justified by the nature of the FCI, where the incorrect multiple-choice answers are based on common misconceptions such as "motion implies active force", "velocity is proportional to force", etc. (Hestenes et al., 1992).

### Network (alluvial) flows to understand pre- to post-test choices

To communicate how these misconceptions change over the course of a semester, we can visualise the data using "alluvial diagrams". While originally developed to track the changes in network structure over time (Rosvall & Bergstrom, 2010), alluvial diagrams have grown in popularity for the ease with which they demonstrate the changing distribution of data points between two states. Another way of thinking of them is that they are similar to Sankey diagrams as used in high school physics to portray energy efficiency in electrical circuits (Australian Curriculum and Assessment Reporting Authority [ACARA], n.d.). They convey how a group will change in a longitudinal type of study from one state to the next, such as with the choices on a pre-test which is then repeated at the post-test stage (repeated measures); for example see Figure 3 for Q4 of the FCI. Such methods appear to have been first used in PER by Yasuda et al. (2023) following on from work in chemistry (Atkinson & Bretz, 2021) and neurophysiology (Doherty et al., 2023)

To appreciate the value of alluvial diagrams (in the context of pre-/post-test comparison studies), consider the intuitive way in which alluvial diagrams are structured: as two vertical axes in parallel (representing the "pre-" and the "post-"state) to which stratum (variables) are assigned, e.g., see Figure 3. Data points are represented with each stratum (block) on each axis as alluviam, with the height of a stratum representing the size of a cluster. Each alluviam is a single alluvial fan (or x-spline) that spans the entire graph. Flows then connect one axis to another. In this way, alluvial diagrams demonstrate the dynamics of the changes between the two phases of the investigation without relying on the statistical prowess of the reader. Since research in physics andragogy, at its core, aims to improve the curriculum for the benefit of students, we must prioritise the communication of our findings clearly and intuitively in order to reach our ultimate audience: educators and policymakers. In the case of this study, we have used the R program and the `ggalluvial` package (Brunson, 2020) to generate the alluvial diagrams.

### Participants, their educational contexts, and the FCI deployment

Traditionally, the FCI is administered as a strictly closed-book 30-minute test held at the beginning of the first semester of students' first university-level physics course (the "pre-test"), to provide instructors with an indication of their students' baseline mechanics skills. At the end of the semester the same FCI test can be administered again (as a "post-test"). Since we track individual answers from students via their student numbers, we can identify "paired data": if a student volunteered to participate in both the pre- and the post-test, we have two sets of "paired" responses that we can compare. As explained earlier, quantities such as the Hake gain are then calculated, to evaluate the effectiveness of instruction over the course of the semester. For both the pre- and post- test, students are not expected to prepare for the assessment; they are not forewarned about it and, upon its completion, the test is not reviewed in class nor do students receive feedback on their attempts.

In this study, we examine the results from participants who took part in our 2022 deployment of the FCI at UJ. These were volunteers from the BEng. programme and the BPhys. programme who sat the FCI pre- and post-test in February and May 2022, respectively. The FCI was deployed via the internal online "learning management system" (LMS), "Blackboard" (Blackboard, 2016), as a 30-minute optional (i.e. zero-credit) review quiz.[6] For the pre-test, results from $N = 176$ BPhys. and $N = 105$ BEng. students were collected; $N = 94$ BPhys. and $N = 230$ BEng. students answered the post-test, see Table 4. Note that we did not include the BPhys. (trad.) participants as the cohorts were considerably smaller (i.e. 28 BPhys. (trad.) participants in the pre-test; see Chrysostomou et al. (2024) for details).

As discussed in the introduction, both the BPhys. and the BEng. programmes are four-year degrees, with a score ≥60% in the final year of high school Mathematics serving as a common prerequisite for the courses (see Table 1). The first introductory physics course for each of these programmes ("Physics 1A1E" and "Engineering Physics 1A", respectively) is a semester-long classical mechanics class that focuses on teaching students "the conceptual foundation for the laws, principles, and methods used in elementary mechanics" (Faculty of Science, 2022); as such, both cohorts should be equipped with the conceptual knowledge and skills to engage successfully with the FCI by the end of the semester. To understand the context more fully, we shall elaborate on the structure of the BPhys. and BEng. programmes (see Table 3 for further details). With this is mind, we also observe that both Physics 1A1E and Engineering Physics 1A fall under the auspices of the Faculty of Science and, specifically, the Department of Physics: as such, despite its emphasis on application, Engineering Physics 1A is taught by physicists rather than by engineers.

The BPhys. and the BEng. degrees share several foundational courses. The similarity in the course structures can be seen in Table 3, where the first semester – the period during which our study takes place – is near identical and covers very similar material. Both programmes begin with "Mathematics 1A1E" and "Engineering Mathematics 1A", covering essential algebra, trigonometry, and introductory calculus; in parallel, both programmes teach classical mechanics in "Physics 1A1E" and "Engineering Physics 1A", respectively. While the "Engineering Physics 1A" course advances at a faster pace than "Physics 1A1E" (covering also some basic thermodynamics in the first semester), both focus on introducing fundamental mechanics and Newton's laws. Both programmes incorporate mandatory courses that introduce students to "professional and technical techniques and standards" (Faculty of Engineering and the Built Environment, 2022) relevant to their chosen career path: specifically, "Computer Competence I" and "Language for Science" for BPhys. students, but "Introduction to Engineering Design 1A" and "Project Communication 1B" for BEng. students. As such, the BPhys. students are being trained to become physicists; the

[6]The test was made available though the Blackboard pages of their first university-level physics courses, i.e. the modules "Engineering Physics 1A" (PHYE0A1) and "Introductory Physics 1AE" (PHY1EA1), respectively, for BEng. and BPhys. students (Faculty of Science, 2022) during the week-long deployment periods specified in Table 4.

BEng. students are being trained for a career in engineering. The distinction we wish to highlight in our comparative study of the two is the different emphases placed on these concepts, based on the training students undergo , and how these influence the DMs exposed by the FCI test. In other words, while these shared courses provide a strong analytical foundation (ensuring that students in both programmes develop an appreciation of the theoretical mathematical and physical sciences), each of the BEng. courses has as a primary goal the application of these theoretical concepts in engineering contexts. We explore in this work how this emphasis on application affects the students' answers of the FCI and what this reveals about their conceptual uptake.

**Table 3**

*A comparison of the course structure of the first two years of BEng. (column 2) and BPhys. (column 3), where the fourth column demonstrates how the latter transitions into the three-year BPhys. (trad.) degree after semester 3. As a tangible example, we consider the course structure for a BEng. in Electrical and Electronic Engineering and a BPhys./BPhys. (trad.) with a Chemistry minor.*

| Sem. | BEng. | BPhys. | BPhys. (trad.) |
|---|---|---|---|
| | Engineering Physics 1A | Physics 1A1E | |
| | Engineering Mathematics 1A | Mathematics 1A1E | |
| 1 | Applied Mathematics 1A (Eng.) | | |
| | Introduction to Engineering Design 1A | Language for Science | |
| | Chemistry 1A | Chemistry 1A1E | |
| | Electrical Engineering Methods 1A | Computer Competence I | |
| | Engineering Physics 1B | Physics 1A2E | |
| | Engineering Mathematics 1B | Mathematics 1A2E | |
| 2 | Applied Mathematics 1B (Eng.) | | |
| | Project Communication 1B | Language for Science | |
| | Electrotechnics 1B | Chemistry 1A2E | |
| | Engineering Physics 2A | Physics 1A3E | Physics S1A |
| | Engineering Mathematics 2A1 & 2A2 | Mathematics 1A1E | Mathematics 1A |
| 3 | Applied Mathematics 2A (Eng.) | | |
| | Electrotechnics 2A | Computer Science IA | ⟨Elective⟩ IA |
| | Modelling 2A | Chemistry 1A3E | Chemistry 1A |
| | Electrical Projects 2A | | |
| | Engineering Physics 2B | Physics S1B | Physics S1B |
| | Engineering Mathematics 2B1 & 2B2 | Mathematics 1B | Mathematics 1B |
| 4 | Applied Mathematics 2B (Eng.) | | |
| | Electrotechnics 2B | Computer Science IB | ⟨Elective⟩ IB |
| | Engineering Economics and Practice 2B | | |
| | Science of Materials 2B | | |
| | Thermodynamics 2B | Chemistry 1B | Chemistry 1B |

As a final comment on the participants involved in this study, we note that the UJ student body is highly representative of the diverse ethnic, cultural, and linguistic backgrounds of South Africa – a country with 12 official languages. Moreover, a sizeable percentage of these students are dependent on government support such as the National Financial Aid Scheme (NSFAS) in order to finance their tertiary education. This socioeco-nomic context has been discussed in other works (Carleschi et al., 2022; Chrysostomou et al., 2024); for detailed demographics of the student body at UJ, see (Faculty of Engineering and the Built Environment, 2022; Faculty of Science, 2022).

**Table 4**

*Physical Science (BPhys.) and Engineering (BEng.) students (see Table 1) and FCI testing details for the first-year participants involved in the* 2022 *FCI pre- and post-test.*

| | Cohort | Deployment period | Resp. / class | Mean (%) |
|---|---|---|---|---|
| Pre-test: | BPhys. | 21/02 - 25/02 | 176/250 | 26.6 |
| | BEng. | 24/02 - 28/02 | 105/500 | 34.7 |
| Post-test: | BPhys. | 24/05 - 30/05 | 94/250 | 31.8 |
| | BEng. | 24/05 - 30/05 | 230/500 | 45.6 |

**Table 5**

*Pre- and post test comparisons for $N$ students, along with the means ($\bar{x}$) % correct answers and %nDMs and Hake gains, for the* 2022 *Engineering (BEng.) and Physical Science (BPhys.) cohorts, participant size $N$; standard deviation in brackets: (SD).*

| Cohort | $N$ | $\bar{x}_{\%Corr}$ (SD) | $\bar{x}_{\%nDM}$ (SD) | Gain |
|---|---|---|---|---|
| Pre Eng. | 105 | 34.7 (14.3) | 51.9 (15.1) | - |
| Post Eng. | 230 | 45.6 (15.1) | 52.9 (13.3) | - |
| Eng. Paired | 54 | - | - | 22.2 |
| Pre Phys. | 176 | 26.6 (12.9) | 49.5 (13.6) | - |
| Post Phys. | 94 | 31.8 (15.6) | 51.2 (15.5) | - |
| Phys. Paired | 70 | - | - | 8.5 |

# Results

## Hake gain

The main results from the "paired" (pre- to post-test) data are shown in Table 5. Note that in the pre- to post-test comparison (Table 5), we have kept all NA attempts, as these will be significant for the alluvial method we discuss shortly. Such NA attempts can lead to lower scores on gains, where typically data cleaning might suggest we drop students who answered less than 20% of all the questions, e.g., see Fazio and Battaglia (2019). Furthermore, and as we mentioned earlier, the definition of Hake gain (Hake, 1998) is not strictly mathematically rigorous (Naylor et al., 2024; Nitta & Aiba, 2019) and hence there is some tension in how to interpret Hake gain anyway. However, we can use Table 5 not just for gains but also to look at means for both cohorts for a comparison with the DMs approach, which also have mean correct and incorrect scores presented.

## DMs for Engineering and Physical Science students

In this section, we present our results for DMs, beginning with Table 5, where as well as the gains, we have also presented results for the mean % score for correct and normalised DMs (nDM) choices. These results are based on the graphical representation of the two cohorts presented in Figures 1 and 2.

The pre-test comparison of results are shown in Figure 1: on the vertical positive axis, a bar graph of the % correct scores is denoted for each question; on the negative axis, the % nDMs are presented. The mean values of the % correct are shown as horizontal lines in the upper part of the graph and are also shown and are listed in Table 5. In the lower part of the graph, only the % threshold is shown (to avoid clutter) rather than the mean value of the %nDM. This graphical representation is different from the tables presented in Martín-Blas et al. (2010, Table 1) who listed question-by-question the % correct and the %nDM. Although this is useful, we are of the view that a graphical snapshot of the cohorts at the pre- and post-test stages is more practical (particularly for educators and policymakers) (Chrysostomou et al., 2024).

The post-test comparison of results are shown in Figure 2. Quite interestingly, although the mean % correct scores increase (as to be expected with improved student performance and in line with student gains), the % nDMs have persisted – and actually appear to have increased. This is also confirmed in Table 5. Part of the reason for this is by design, as the % correct answer is also included in the normalisation: in our definition of the nDM (see Equation (2)), we subtract the % correct answer from the denominator; larger % correct implies a smaller denominator, which in turn bloats the % nDM. Hence, we expect there to be an increase in the % nDM scores if the % correct scores also increases.[7] This is a useful aspect of nDMs because this means that even if only the post-test stage is administered a researcher is still able to identify which questions have "persistent" misconceptions (Bani-Salameh, 2016a).

An important observation is the issue of persistent misconceptions. As can be seen from Figure 1 as compared to Figure 2, is that the DMs still remain at the post-test phase.

[7]Many of the % nDM questions were close to the threshold ($y = 50\%$) at the pre-test stage and if the percentage correct increases we then will see more questions move above that threshold as nDMs are normalised by the % correct, see Eq. 1.

This is especially clear when we focus on questions: Q3, Q4, Q6, Q10, Q15, Q17, and Q30. These questions, using the conceptual breakdown of (Hestenes et al., 1992), see appendices, are related to concepts of Newton's II and III law, that the sum of vectorial forces is zero and in one case circular motion (though related to how vectors are also handled, Q6). This seems to point to underlying mathematical issues that have "persisted" during the presentation of the course. But given that it is not made clear in this analysis whether the same option (of the five given) was chosen each time, where certain choices would indicate certain conceptual weaknesses, probing the precise choices made becomes a focus of this analysis through the use of alluvial diagrams.

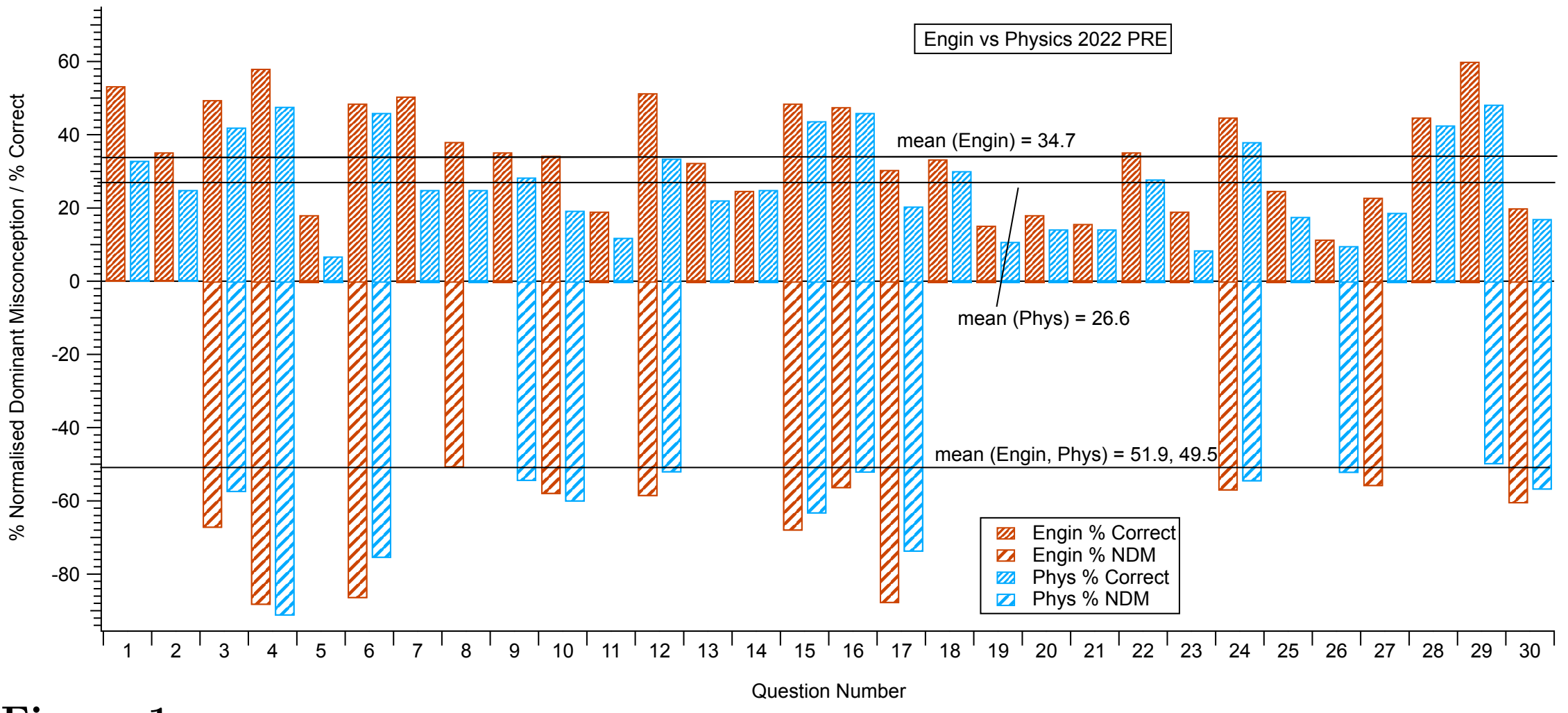

**Figure 1**

Pre-test score breakdown for the 2022 Engineering (BEng.) and Physical Science (BPhys.) cohorts. The positive axis' horizontal black lines correspond to the mean values of % correct for each cohort. On the negative axis, the black line is the cut at $y = 50\%$ set as the threshold.

As a further example of the utility of nDMs, we can compare our results with data from other studies that have also used DMs based in other countries. The data are presented in Table 6, where for comparison we have focused on the questions discussed by Martín-Blas et al. (2010) (who considered Spanish Engineering students) along with data from the KSA (Bani-Salameh, 2016a), at the pre-test stage. The six questions: Q4, Q28, Q19, Q2, Q9, and Q14 given in Table 6 are grouped based on their conceptual structure, see A1. We can see similarities and differences in the % nDM scores between the various groups. For example, all groups and countries follow the same trend for questions: Q4, Q28, Q19, and Q9. On the other hand, Q2 and Q14 have some difference between UJ students and the other groups / countries. In the UJ case, and given that three years of data also give similar results (see Table 6), it appears that South African students did not have dominant misconceptions for these two questions compared to the other three groups outside of South Africa. As can be seen from Figure 1, for Q2 and Q14 both cohorts have relatively low % correct scores which might imply lower % nDMs, see Equation 2. However, it could also be that other incorrect choices were chosen and that there was no dominant incorrect choice – we discuss further in the next section.

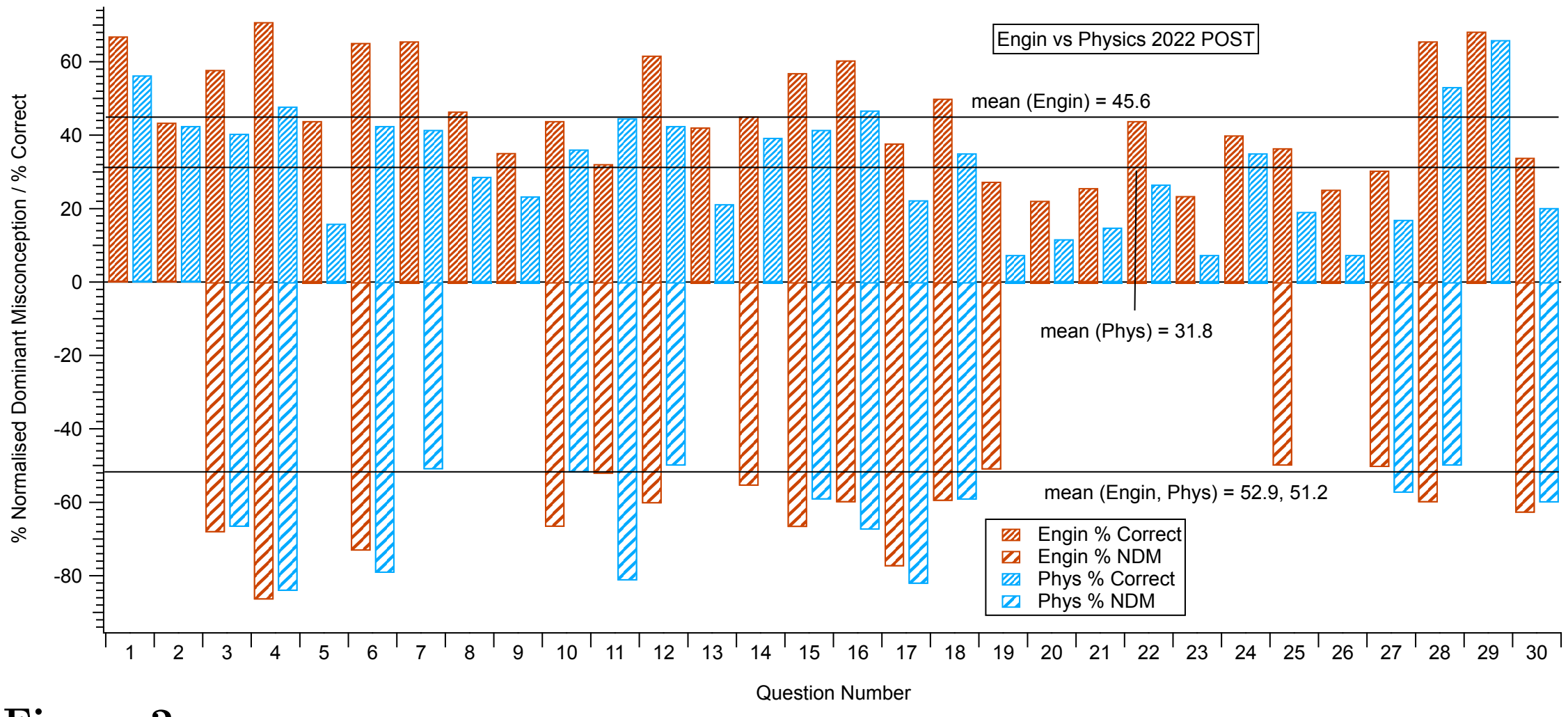

**Figure 2**

*Post-test score breakdown for the* 2022 *Engineering (BEng.) and Physical Science (BPhys.) cohorts; same labelling conventions as in Figure 1.*

**Table 6**

*A comparison of pre-test %nDMs from Martín-Blas et al. (2010) (Groups 1 and 2) and Bani-Salameh (2016b) (rounded to nearest whole number) with the 2022 Eng. and Phys. cohorts. The three year UJ Average (2020 to 2022) is also shown which comprises of* $n = 804$ *students in total, Chrysostomou et al. (2024).*

| | Q4 (%) | Q28 (%) | Q19 (%) | Q2 (%) | Q9 (%) | Q14 (%) |
|---|---|---|---|---|---|---|
| Group 1 (MB) | 95 | 68 | 44 | 89* | 56 | 96 |
| Group 2 (MB) | 96 | 64 | 58 | 76* | 65 | 86 |
| Bani-Salameh | 73 | 56 | 67 | 58 | 49 | 90 |
| Eng. 2022 | 88 | 56 | 46 | 35 | 51 | 49 |
| Phys. 2022 | 91 | 49 | 40 | 27 | 45 | 33 |
| UJ 3-yr. av. | 92 | 54 | 44 | 32 | 52 | 48 |

*These values were presented as 68 and 64 for Group 1 and 2, respectively, in Bani-Salameh (2016b) and Chrysostomou et al. (2024). We have confirmed that the values are as above (Martín-Blas et al., 2010, Table 1).

In the UJ context, using Tables 5 and 6, as well as Figures 1 and 2, we can also compare % correct choices for BEng. and BPhys. students. Although the BEng. students in general had larger % correct choices in the pre- and post-tests, see Figure 5, at the same time many of the %DMs (or % incorrect responses) were larger. As we mentioned, this is also connected to the fact that the nDMs are normalised from Equation 2 by the % of correct responses and lead to what we call persistent misconceptions. This indicates possibly that the conceptually "stronger" cohort was also more likely to choose the next best option, whereas the BPhys. cohort (who also had an easier entrance score, see Table 4), are more likely to select their answers arbitrarily. To confirm if this is indeed the case, we analyse the same set of six questions through alluvial flow diagrams.

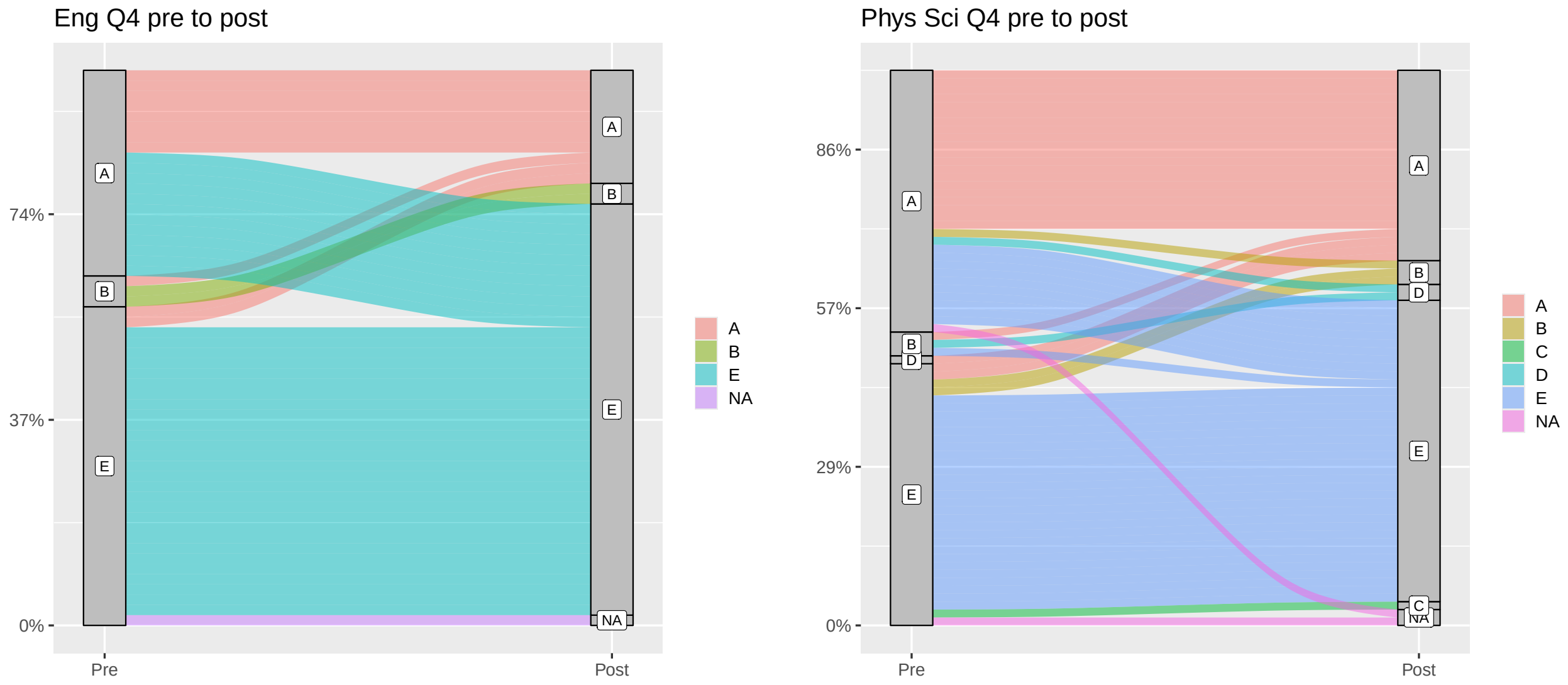

**Figure 3**

*Q4 for Engineering (BEng.) and Physical Science (BPhys.) students. On the vertical axis we have the number in each cohort, $N = 54$ and $N = 70$, respectively (both axes are; however, scaled to the same % magnitude). For Physical Science Q4, choices C and NA were not present at the pre-test stage, but were at the post-test stage and hence the out-of-sync stratum labels on the alluvial. Ans: E*

## Alluvial diagrams for Engineering and Physical Science cohorts

As we discussed earlier, we wish to track students' choices in some way for both cohorts, to confirm if the results indicated by the nDMs are indeed persistent misconceptions. At the same time, the normalisation of such nDMs could be a factor, so in this section we will utilise alluvial diagrams as a visual means of analysing the particular choices made by students for a given question, from the pre- to post-test stages. To give some relevant context we make a comparison of Q4 through to Q14, see Table 6 for DMs, which are those considered in other countries as well. The alluvial diagrams for these questions are presented in Figures 3, 4, 5, 7, 6, and 8.

For example in Figure 3 we can compare the BEng. and BPhys. students on the left and right panes, respectively. The correct answer to Q4 is choice E, and we see that from the pre- to post-test stages the % in each group has increased. We also notice that there is a greater % of BPhys. students who have not only selected answers other than the correct choice at the pre-test stage, but have also made more jumps to other choices at the post-test stage. In contrast, for the Engineering students, a large % updated their choices

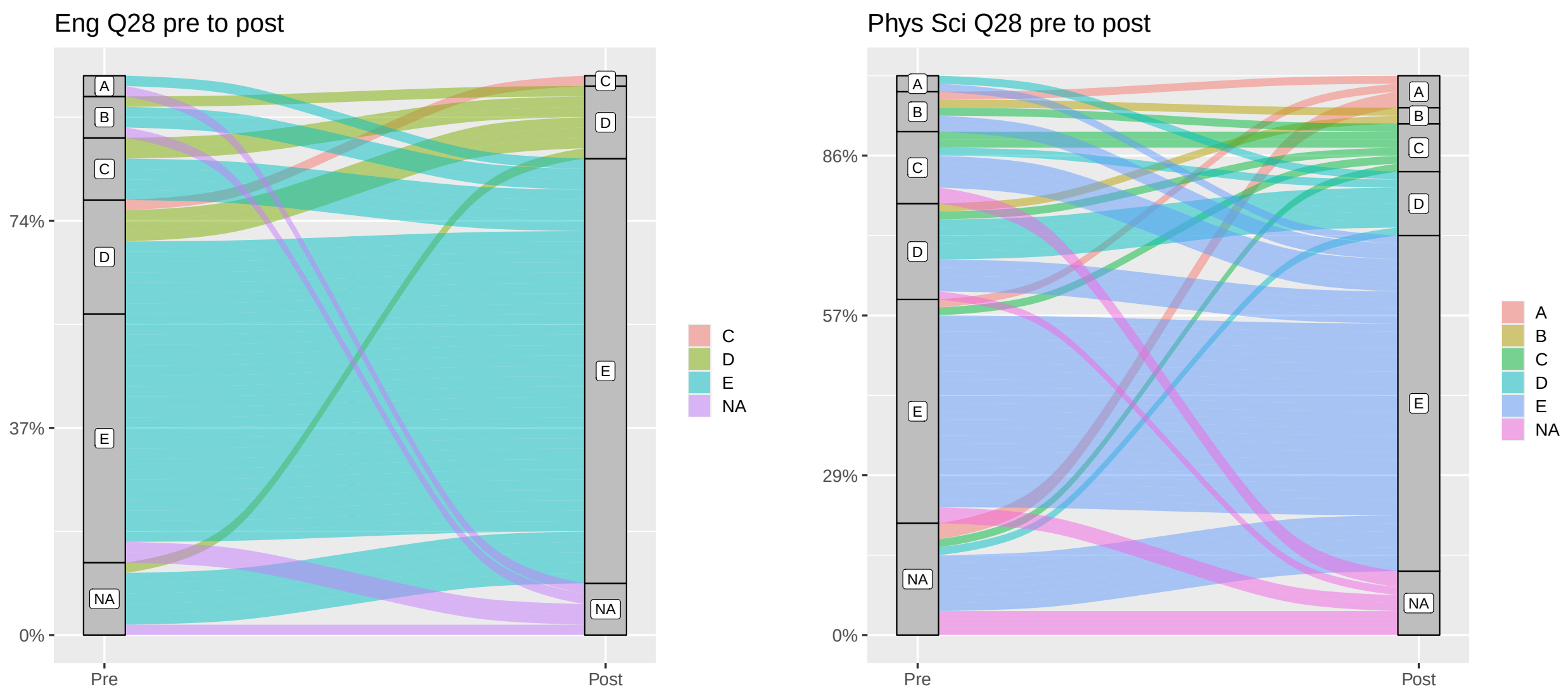

**Figure 4**

*Q28 for Engineering (BEng.) and Physical Science (BPhys.) students with the same nomenclature as before. Ans: E*

from choice A to choice E (the correct choice); although some BPhys. students did this, there were many more who updated their choices incorrectly. Alluvial diagrams, also allow us to follow "Non–Answer" or NA responses, where as opposed to Q4, Q28 has more of these.

A slightly different pattern emerges for Q28, where we see that as before the BEng. students have a larger group with the correct choice at the pre- and post-test stages as compared to Q4. However, we see there are more "jumps", where the student has updated their choice from the pre- to post-test stages. Whilst there appears to be more jumps for the BPhys. as compared to the BEng. students both groups do have these updated choices. Q19 and Q9 follow similar patterns with even more jumps between pre- and post-test choices. Again we see better scores for the BEng. students. However, for Q9 we see a drop in correct choices, which can also be discerned from Figures 1 and 2. A table of % correct choices / %nDMs (Martín-Blas et al., 2010, Table 1) or the use of the bar graphs (Figures 1 and 2) allow a researcher or lecturer to gather a snapshot of information with the alluvial diagrams giving another way to visualise how % correct choices change from pre- to post-test stages.

We now consider Q2 and Q14 in relation to the interesting differences we found in Table 6 in comparison to Spain and the KSA. In our case we have found that nDMs for these two questions were much lower than the results found in Spain (Martín-Blas et al., 2010) and the KSA (Bani-Salameh, 2016b). Although the difference may be due simply to the random nature of different test groups, of which there are 30 different questions in the FCI, it is still interesting to try and use alluvial diagrams to understand why the UJ groups scored differently. From the alluvial plot for Q2 in Figure 7, we can see that the correct answer choice (A) has the largest % and then B and D both are of the same magnitude as the largest % incorrect. This is to be contrasted with Q4, Q28, Q19 and Q9, where there is a clear dominant incorrect choice. In the case of Q2, this is more pronounced for the BPhys. students, see Figure 7. We also see a similar pattern in Q14 (Figure 8), where besides the correct answer choice, there are no real dominant incorrect choices at the pre-test stage. It is interesting to note that at the post-test stage, Q14 develops a %nDM above the threshold of 50% (see Figure 2), as can also be seen from the alluvial in Figure 8 for the Engineering students. For the BPhys. students, there is an observed gain for questions: Q2 and Q14, but there is no dominant incorrect choice.

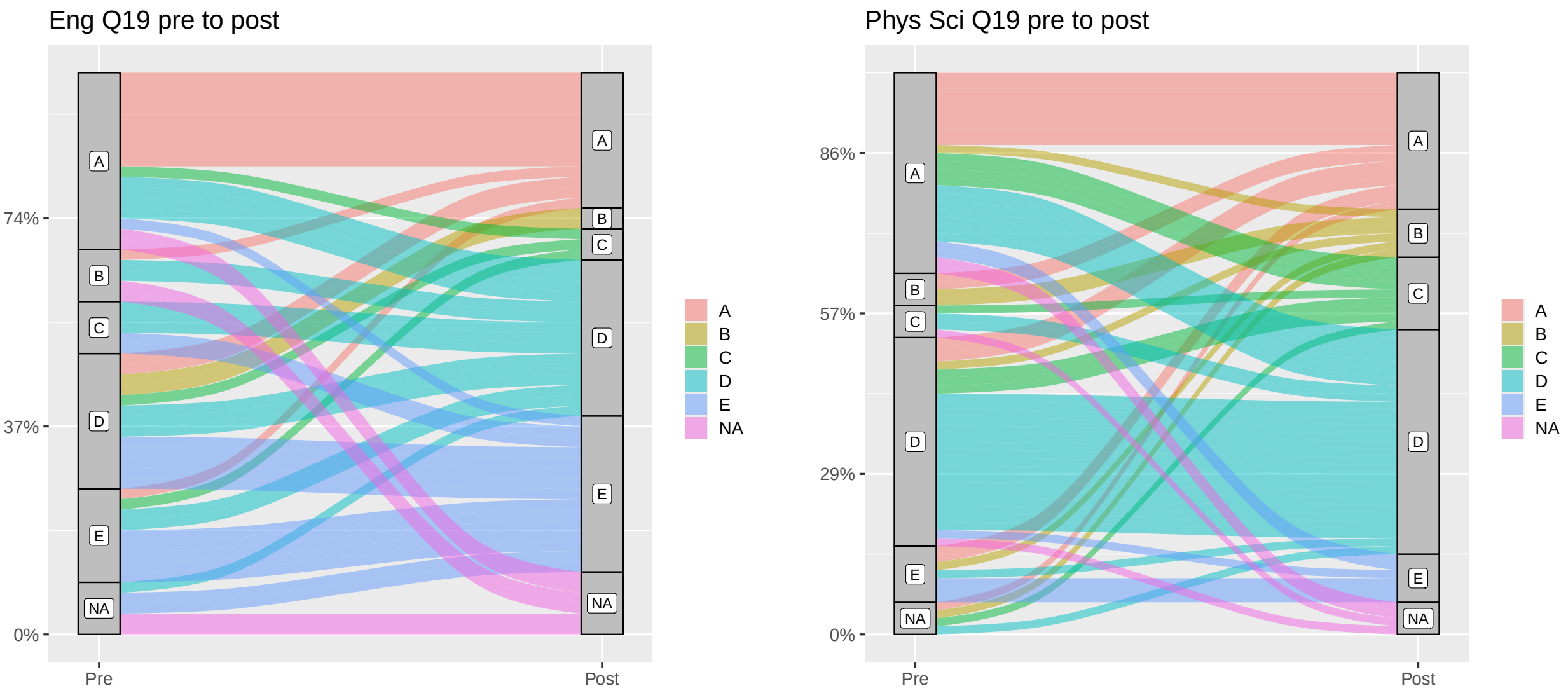

**Figure 5**

*Q19 for Engineering (BEng.) and Physical Science (BPhys.) students with the same nomenclature as before. Ans: E*

## Conclusion & implications

In this work we have discussed how the idea of DMs can be used to give the "teacher" the ability to make "informed choices" and adapt to the conceptual learning needs of the class. Previous research on DMs by Bani-Salameh (2016a, 2016b), Bouzid et al. (2022), and Martín-Blas et al. (2010) have elegantly highlighted how DMs can be employed to adapt one's teaching. Our graphical (bar chart) approach to DMs in conjunction with the use of alluvial diagrams (Yasuda et al., 2023) has allowed us to answer our research questions.

In response to our initial research questions, we have found that:

**RQ1–** The extent to which dominant misconceptions compare between Physical Science (BPhys.) and Engineering (BEng.) students at the University of Johannesburg was significant in terms of differences between gains and mean scores for % correct choices. However, normalised DMs remained consistent within both groups.

**RQ2–** Alluvial diagrams confirm the gains and mean score differences between Physical Science and Engineering students, with the former group following a more random

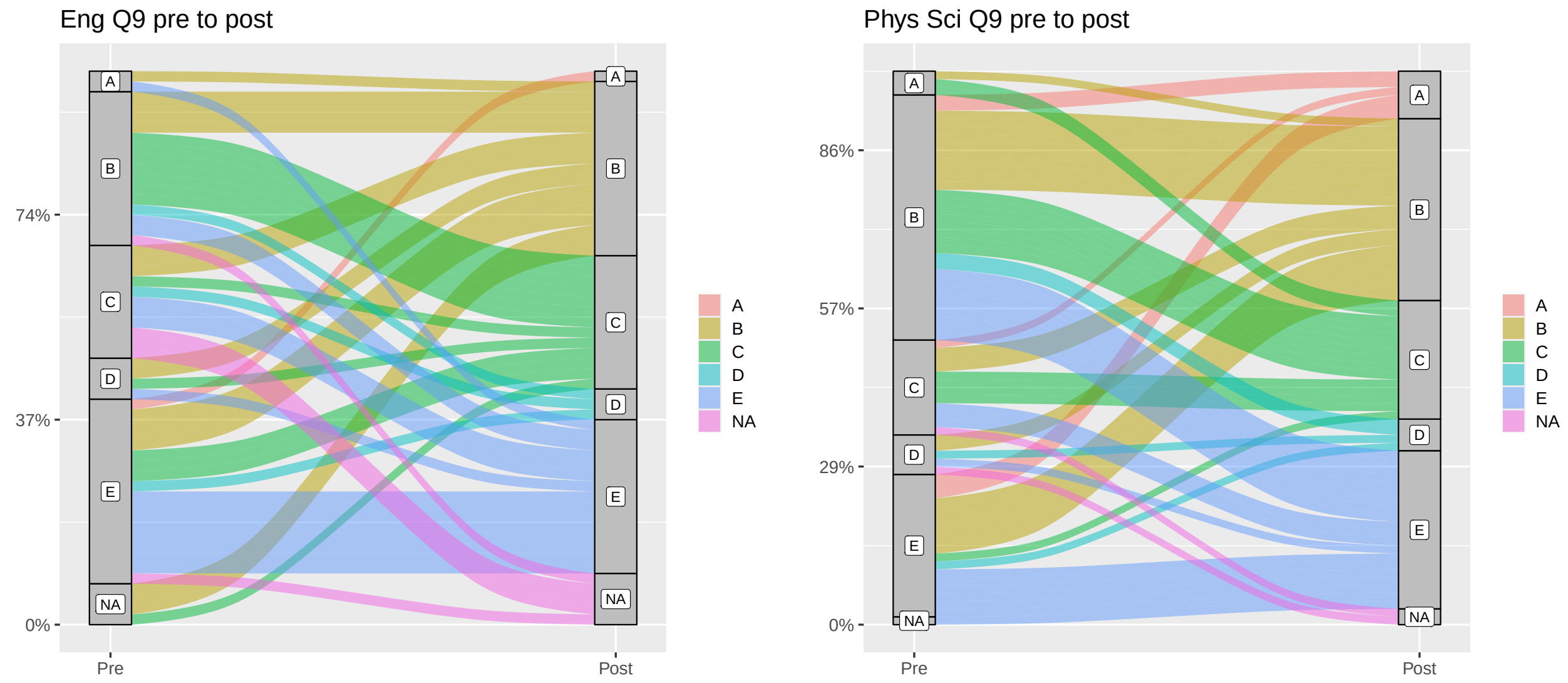

**Figure 6**

*Q9 for Engineering (BEng.) and Physical Science (BPhys.) students with the same nomenclature as before. Ans: E*

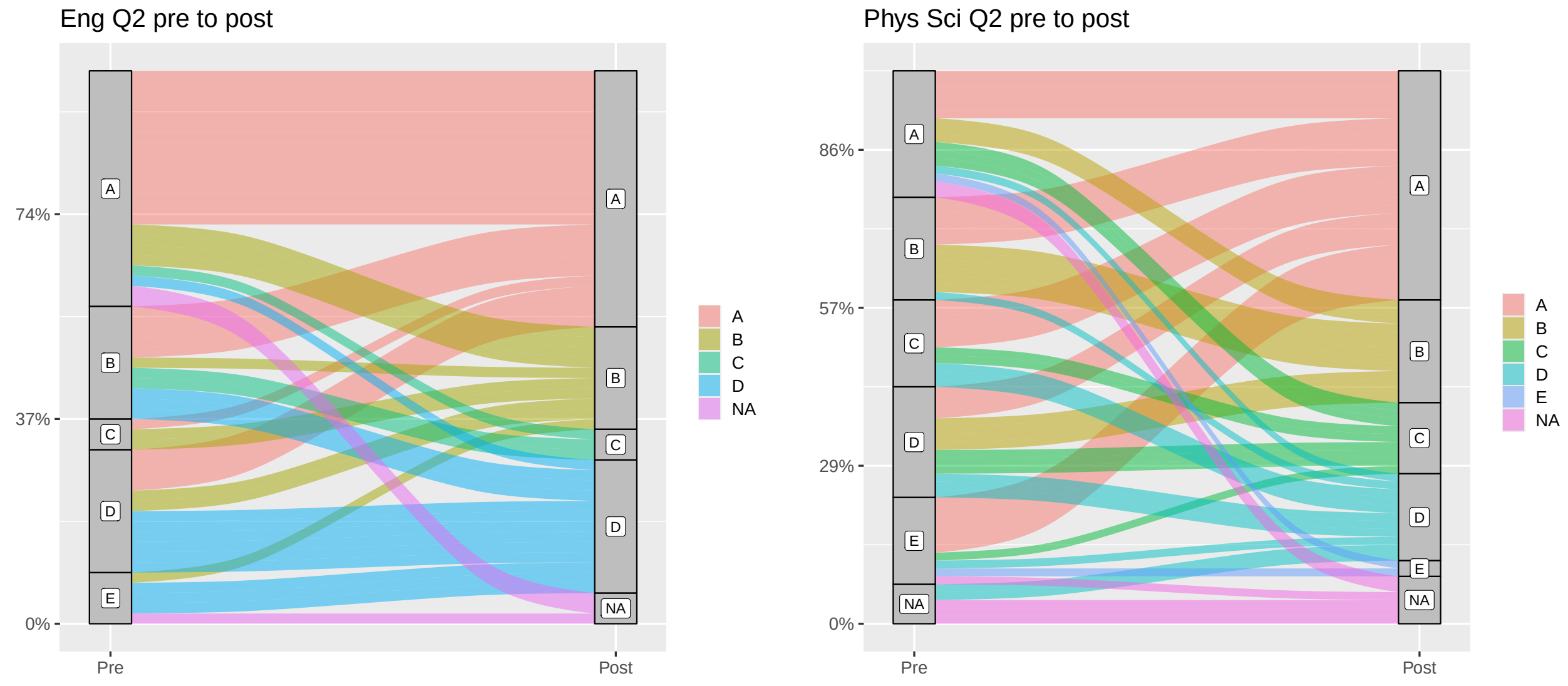

**Figure 7**

*Q2 for Engineering (BEng.) and Physical Science (BPhys.) students with the same nomenclature as before. Ans: A*

(chaotic) flow of pre- to post-test answer choices, for the subset of FCI questions we have looked at: Q4, Q28, Q19, Q2, Q9, and Q14.

In relation to persistent DMs in RQ1, this is partly due to the design of normalised DMs, see Equation 2, and is useful because at either pre- or post-test stages a lecturer can check which questions still have persistent misconceptions. In terms of the pre-test this can

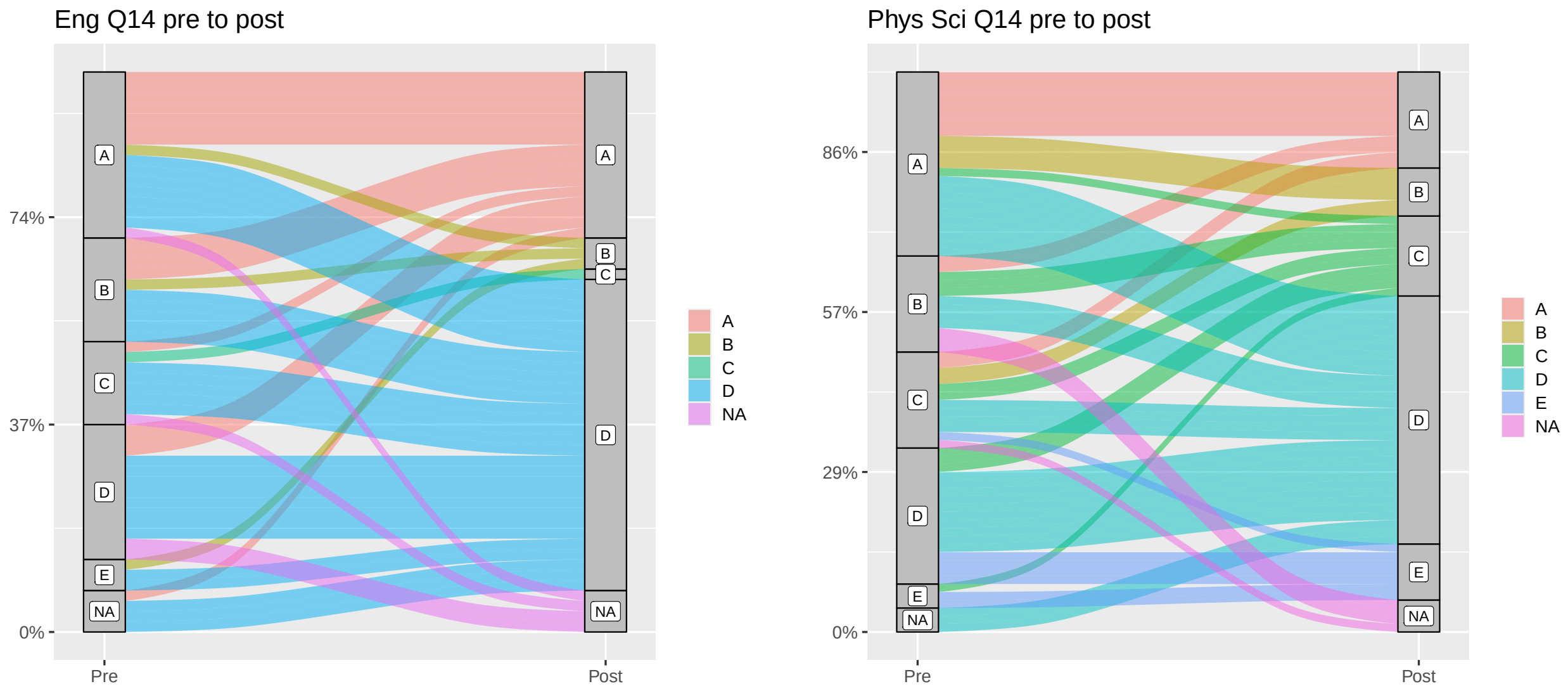

**Figure 8**

*Q14 for Engineering (BEng.) and Physical Science (BPhys.) students with the same nomenclature as before. Ans: D*

update one's andragogy during the teaching semester, and can be checked at the post-test stage (if administered). In terms of the post-test stage this can be used for planning for the next presentation of the course.

In relation to alluvial diagrams in RQ2, we have looked at a subset of questions (Q4, Q28, Q19, Q2, Q9, and Q14) for conciseness and as is standard in the literature (Atkinson & Bretz, 2021; Yasuda et al., 2023). These were the same questions used for comparison with cohorts from other countries, see Table 6. Note though that it is straightforward to generate alluvial diagrams for any of the FCI questions. Aside from being able to easily compare with previous studies, these questions are particularly useful given the cross-section of the concepts they assess. For example, from Table A2, Q2, Q4 and Q28 deal with Newton II and III; whilst Q9 deals with vector addition; Q14 looks at projectile motion; and Q19 looks at velocity versus acceleration. In terms of conceptual coding, see Table A1 (Hestenes et al., 1992), the spectrum of concepts covered by these questions is very wide. As such, when we look at the alluvial diagrams we observe that for the Engineering students they answer in the post-test with almost exclusively either the persistent misconception or the correct answer. However, Physical Science students' responses are more "chaotic", with the set of choices being made in the post-test moving towards almost all options. Note that this behaviour exists for all the questions studied. As such, this points to a more general feature of how these students answered these tests.

## Limitations and future work

There are of course limitations to what we have presented as we have only suggested that the entrance requirements for each of the two student cohorts is a "factor" in terms of FCI performance. Given more data and the use of each students matriculation results it would be interesting and possible to compare the Engineering and Physical Science students FCI scores to each students matriculation result using correlation analyses (Hewagallage et al., 2022) or structural equation modelling (SEM) (Cwik & Singh, 2022). Although the repeated measures (paired) data sample size is quite small (order $N = 50$) it would be possible to look at the various cohorts that we have studied (from 2020 – 2022, (Naylor et al., 2022)) which are of the order $N = 500$ where SEM type analyses of matriculation scores and pre-test scores could be used to see which factors, such as English, Science or Mathematics scores correlate most with pre-test scores and also dominant misconceptions. We leave this for future endeavours.

A further limitation to our study lies in the fact that both Engineering Physics 1A and Physics 1A1E were taught by physicists. As such, even though the Engineering students followed a course that emphasises applications of physics rather than theory, we must acknowledge the bias of the educator towards physics training. If the methods of instruction used by physicists serve as a possible source of student misconception, we would anticipate the same misconceptions to propagate within the engineering cohort. This suggests more collaboration between physicists and engineers in faculty teaching.

Finally, it would also be interesting to look at how cohorts at UJ and in other countries perform on the FCI via interventions directly related to using DMs / alluvial diagrams at the pre-test stage. That is to say using RBAs to inform our teaching with specific interventions guided by the results from DMs and alluvial diagrams such that lecturers update their teaching during the course; rather than in future year's presentations. Such an enterprise would require a whole cross-faculty approach (between Science & Engineering Faculties in the UJ context), or at least the first year academic coordination team to have all lecturers and tutors on board such that all physics cohorts have the same intervention to be equitable to all students, including input from Engineering.

## Acknowledgements

The authors would like to acknowledge open access publishing facilitated by ACU as part of an agreement via the Council of Australian University Librarians. AC acknowledges the support of the National Research Foundation (NRF) of South Africa and Department of Science and Innovation through the SA-CERN programme, as well as that of a Campus France scholarship and a research grant from the L'Oréal-UNESCO's *For Women in Science* Programme. She is now supported by the Initiative Physique des Infinis (IPI), a research training program of the Idex SUPER at Sorbonne Université. ASC is supported in part by the NRF. WN is supported by a Faculty of Education and Arts Grant, Project Code No: 50-905300-111. The authors express their sincere thanks to the participating students and the lecturers at the Physics Department, UJ.

## Note on contributors

Anna Chrysostomou is a postdoctoral fellow at the Laboratoire de Physique Théorique et des Hautes Énergies (Sorbonne Université). She obtained her PhD degree in mid-2024 through the University of Johannesburg and the Université Claude Bernard Lyon-1. Her research lies in high-energy theoretical physics, with a specific focus on gravitational physics and its intersection with particle physics and cosmology.

Alan Cornell is a professor of theoretical physics at the University of Johannesburg. He studied in Australia and has worked in Korea, Japan, and France. His research interests are in particle and gravitational physics, and higher education pedagogy.

Wade Naylor (MInstP) worked formerly as a theoretical physicist for over 15 years including as an associate professor at Osaka University. After a career change to become a high school physics teacher, Wade is now a lecturer in STEM education at ACU. Wade's research interests lie in Physics Education Research (PER) and how to understand students' 'misconceptions' in physics and STEM using CIs. He is also a visiting senior researcher at the Department of Physics, University of Johannesburg, South Africa.

## Ethical statement

The authors confirm that all participants in the study gave consent to participate in the study, and consent for the results to be published was obtained from participants for all research involving human subjects. The request to conduct this study (in accordance with the principles embodied in the Declaration of Helsinki and local statutory requirements) for the project entitled 'Identifying conceptual issues in first year physics curricula,' was granted by the ethics committee at the UJ: Project Code No. 2023-02-03/CornellR1. This includes studies done on teaching methods.

## Appendix

### Supplemental: Taxonomy of Concepts for FCI Questions

For completeness, we present two tables to help describe the taxonomy of conceptual understanding and hence possible 'misconception' issues for each question in the FCI. These tables are based on the work by Bani-Salameh (2016a), Bayraktar (2009), and Hestenes et al. (1992) and Chrysostomou et al. (2024).

Using Table A1, the following table, Table A2, gives information about the breakdown of each question response conceptually and is similar in vein to that presented in the work of Bani-Salameh (2016a). Note that not every answer response ($A, B, C, D, E$) has a conceptual code.

**Table A1**

*Conceptual coding for the FCI as based originally in Hestenes et al. (1992).*

| Code | Misconception (Pre-conception) |
|---|---|
| K1 | Position-velocity undiscriminated |
| K2 | Velocity-acceleration undiscriminated |
| K3 | Non-vectorial velocity composition |
| K4 | Ego-centered reference frame |
| I1 | Impetus supplied by 'hit' |
| I2 | Loss/recovery of original impetus |
| I3 | Impetus dissipation |
| I4 | Gradual/delayed impetus build-up |
| I5 | Circular impetus |
| AF1 | Only active agents exert forces |
| AF2 | Motion implies active force |
| AF3 | No motion implies no force |
| AF4 | Velocity proportional to applied force |
| AF5 | Acceleration implies increasing force |
| AF6 | Force causes acceleration to terminal velocity |
| AF7 | Active force wears out |
| AR1 | Greater mass implies greater force |
| AR2 | Most active agent produces greatest force |
| CI1 | Largest force determines motion |
| CI2 | Force compromise determines motion |
| CI3 | Last force to act determines motion |
| CF | Centrifugal force |
| Ob | Obstacles exert no force |
| R1 | Mass makes things stop |
| R2 | Motion when force overcomes resistance |
| R3 | Resistance opposes force/impetus |
| G1 | Air pressure-assisted gravity |
| G2 | Gravity intrinsic to mass |
| G3 | Heavier objects fall faster |
| G4 | Gravity increases as objects fall |
| G5 | Gravity acts after impetus wears down |

**Table A2**

*Conceptual breakdown of FCI questions $1-30$. These concepts are grouped roughly into: Newton's three laws, circular motion, s vs t graphs, projectile motion, vector addition and $\Sigma F_{net}=0$. The coding of each question response (from A to E) are given in Table A1.*

| | Law/concept | A | B | C | D | E |
|---|---|---|---|---|---|---|
| 1 | Newton II | G3 | | | | |
| 2 | Newton II | | G3 | | G3 | |
| 3 | Newton II | AF6 | AF5, G4 | | G2 | G1 |
| 4 | Newton II, III | AR1 | | Ob | AR1 | |
| 5 | circ. motion | Ob | | I1, I5, AF2 | I1, I5, AF2 | I1, I5, AF2, CF |
| 6 | circ. motion | I5 | | CF | CI2, CF | CF |
| 7 | circ. motion | I5 | | CI2, CF | I2, I5, CF | CF |
| 8 | s vs t | CI3 | | I2 | I4 | I2 |
| 9 | vec. add. | | CI3 | K3 | | |
| 10 | $\Sigma F_{net}=0$ | | I4 | | I4 | |
| 11 | $\Sigma F_{net}=0$ | Ob, G1 | I1, Ob | I1 | | G2 |
| 12 | proj. motion | CI2 | | I3 | I3, G5 | |
| 13 | Newton II | I3 | I3, G4, G5 | I3 | | G2 |
| 14 | proj. motion | K4 | K4 | CI2 | | I3, G5 |
| 15 | Newton III | | AR1 | AR2 | AF1 | Ob |
| 16 | Newton III | | AR1 | AR2 | AF1 | Ob |
| 17 | $\Sigma F_{net}=0$ | CI1 | | | CI1, G1 | AF1 |
| 18 | circ. motion | AF1, Ob | | I5 | I5 | CF |
| 19 | vel. vs acc. | K2 | K1 | K1 | K1 | |
| 20 | vel. vs acc. | | K2 | K2 | | |
| 21 | Newton II, s vs t | I2 | CI3 | CI2 | I4 | |
| 22 | Newton II | AF4 | | AF7 | AF6 | AF7 |
| 23 | s vs t, $\Sigma F_{net}=0$ | I2 | | CI3 | I2, I3 | I4 |
| 24 | $\Sigma F_{net}=0$ | | | I3 | | I3 |
| 25 | $\Sigma F_{net}=0$ | R2 | R2 | | R2 | CI1 |
| 26 | Newton II | AF4 | R2, R3 | I4 | AF6 | |
| 27 | Newton II | AF2, R1 | I3, R1 | | I1 | I4 |
| 28 | Newton III | | AF1 | | AR1, AR2 | |
| 29 | $\Sigma F_{net}=0$ | Ob | | G1, G2 | G1 | AF3 |
| 30 | Newton II | AF1 | I1 | | I1 | I1 |